\documentclass[sigconf]{acmart}
\renewcommand\footnotetextcopyrightpermission[1]{}

\usepackage[english]{babel}
\usepackage{blindtext}
\usepackage{algpseudocode}
\usepackage{booktabs}
\usepackage{pifont}
\usepackage{utfsym}
\usepackage{enumitem}

\newcommand{\ours}{{WebSplatter}\xspace}

\AtBeginDocument{%
  \providecommand\BibTeX{{%
    \normalfont B\kern-0.5em{\scshape i\kern-0.25em b}\kern-0.8em\TeX}}}

\acmISBN{978-1-4503-XXXX-X/18/06}

\begin{document}

\title{\ours: Enabling Cross-Device Efficient Gaussian Splatting in Web Browsers via WebGPU}

\author{Yudong Han}
\email{hanyd@pku.edu.cn}
\affiliation{%
  \institution{Institute for Artificial Intelligence, Peking University}
  \country{China}
}

\author{Chao Xu}
\email{eric.xc@alibaba-inc.com}
\affiliation{%
  \institution{Tongyi Lab, Alibaba Group}
  \country{China}
}

\author{Xiaodan Ye}
\email{doris.yxd@alibaba-inc.com}
\affiliation{%
  \institution{Tongyi Lab, Alibaba Group}
  \country{China}
}

\author{Weichen Bi}
\email{biweichen@pku.edu.cn}
\affiliation{%
  \institution{Institute for Artificial Intelligence, Peking University}
  \country{China}
}

\author{Zilong Dong}
\email{list.dzl@alibaba-inc.com}
\affiliation{%
  \institution{Tongyi Lab, Alibaba Group}
  \country{China}
}

\author{Yun Ma}
\email{mayun@pku.edu.cn}
\affiliation{%
  \institution{Institute for Artificial Intelligence, Peking University}
  \country{China}
}

\begin{abstract}

We present WebSplatter, an end-to-end GPU rendering pipeline for the heterogeneous web ecosystem. Unlike naive ports, WebSplatter introduces a wait-free hierarchical radix sort that circumvents the lack of global atomics in WebGPU, ensuring deterministic execution across diverse hardware. Furthermore, we propose an opacity-aware geometry culling stage that dynamically prunes splats before rasterization, significantly reducing overdraw and peak memory footprint. Evaluation demonstrates that WebSplatter consistently achieves 1.2$\times$ to 4.5$\times$ speedups over state-of-the-art web viewers.

\end{abstract}




\maketitle

\pagestyle{plain} 
\thispagestyle{plain} 

\section{Introduction}

The web has become one of the dominant platforms for interactive 3D experiences, driven by its accessibility, zero-install deployment, and cross-platform sharing~\cite{dominant1,dominant2}. As browser technologies and graphics APIs continue to evolve, rendering complex, large-scale scenes directly on the web has shifted from a niche capability to a mainstream demand. To meet this need for both high fidelity and real-time performance, 3D Gaussian Splatting (3DGS)~\cite{kerbl20233d} has emerged as a revolutionary technique, rapidly gaining traction within the web ecosystem.
Its differentiable nature makes it particularly appealing for generative and reconstruction tasks, where models learn to synthesize or reconstruct 3D scenes from images, videos, or text prompts. Therefore, 3DGS is rapidly becoming a building block in the emerging ecosystem of generative 3D modeling~\cite{tang2023dreamgaussian,zhou2024diffgs,lin2025diffsplat}, supporting applications for high-fidelity reconstruction, neural rendering, text-to-3D generation, and interactive content creation in the browser.
This trend is further amplified by Marble~\cite{Marble2025}, a creative platform recently introduced by Li Fei-Fei’s WorldLabs, in which creators generate entire virtual worlds and export them as Gaussian splats for reuse in downstream projects. This convergence of web-native deployment, real-time rendering, and generative 3D modeling is reshaping the landscape of immersive content creation. By enabling seamless distribution and easy access through the browser, it lowers the barrier to producing and sharing high-quality 3D scenes. As a result, bringing 3DGS to the web is no longer a novelty but an essential and timely research direction~\cite{zhu_rendering_2025, sinha_spectralsplatsviewer_2024}.

Despite the growing interest in 3DGS, building a high-performance and portable renderer for the web is still challenging and unexplored. The difficulty lies in a fundamental mismatch between traditional graphics pipelines and neural rendering. WebGL~\cite{webgl1,webgl2}, the conventional web graphics API, follows a fixed-function design optimized for triangle meshes. In contrast, 3DGS requires intensive, frame-dependent computations, such as depth sorting and view-adaptive color evaluation. Since WebGL lacks general-purpose compute capabilities, these tasks are offloaded to the CPU via JavaScript or WebAssembly, introducing latency and bandwidth overheads that scale poorly with scene complexity~\cite{zhu_rendering_2025}. Consequently, real-time rendering of large-scale 3DGS scenes remains impractical under the WebGL API.

The emergence of WebGPU~\cite{webgpu} offers an opportunity to overcome these limitations by introducing modern GPU features to the web browser. These features include compute shaders, storage buffers, and fine-grained resource control, which enable complex workloads such as sorting and shading to be executed entirely in parallel on the GPU, thereby eliminating most CPU-GPU synchronization overheads. However, recent WebGPU-based 3DGS renders have encountered new challenges. Many of these ports are directly adapted from CUDA or Vulkan code. However, the web is heterogeneous, spanning a wide range of devices from high-end discrete GPUs to mobile SoCs. This diversity often leads to performance degradation or even synchronization failures on certain platforms. As a result, despite the modern GPU capabilities, achieving robust and efficient 3DGS rendering across browsers and devices remains a significant open problem.

To address this gap, we propose \ours \footnote{The source code is available at \url{https://anonymous.4open.science/r/webgs}.}, a WebGPU-based framework for real-time 3DGS. \ours addresses web-specific challenges in WebGPU and leverages a fully GPU compute-render pipeline to overcome the performance and portability limitations of WebGL-based approaches. Designed for heterogeneous hardware and vendor-neutral execution, it enables efficient rendering across desktop and mobile platforms. The main contributions are summarized as follows:

\begin{itemize}[leftmargin=1.5em]
    \item \textbf{An end-to-end WebGPU framework for 3D Gaussian Splatting.} We present \ours, the first framework that bridges the gap between native GPU capabilities and the sandboxed web environment. Unlike direct ports of native algorithms, which fail due to the lack of low-level thread interlocks in browsers, \ours employs a wait-free, multi-stage parallel pipeline designed specifically to guarantee synchronization and deadlock-free execution on the Web.

    \item \textbf{Optimization for cross-platform WebGPU backends.} We propose a set of techniques that ensure stable and efficient performance across diverse devices. A \emph{wait-free radix sort} eliminates inter-workgroup spin waiting and guarantees deadlock-free execution on GPUs from different vendors. Screen-space culling and opacity-driven quad sizing further reduce the overall computational workload and improve runtime efficiency.

    \item \textbf{Comprehensive Evaluation on Performance.} We evaluate \ours across diverse hardware platforms. The framework achieves speedups ranging from $1.18\times$ to $4.5\times$ over state-of-the-art baselines. Furthermore, we demonstrate that \ours reduces peak memory consumption compared to existing WebGPU ports, preventing crashes on memory-constrained devices.
\end{itemize}

The remainder of this paper is organized as follows. Section~\ref{sec:background} introduces the fundamentals of 3DGS and the challenges of implementing it on WebGPU. Section~\ref{sec:design} presents the architecture of \ours. Section~\ref{sec:evaluation} reports evaluation results and performance analyses across multiple hardware platforms. Section~\ref{sec:relatedwork} reviews existing studies on WebGPU, acceleration techniques for Gaussian Splatting, and web-based 3D rendering. Finally, Section~\ref{sec:conclusion} concludes the paper.

\section{Background}\label{sec:background}

\begin{figure*}[t]
\centering
\includegraphics[width=\textwidth]{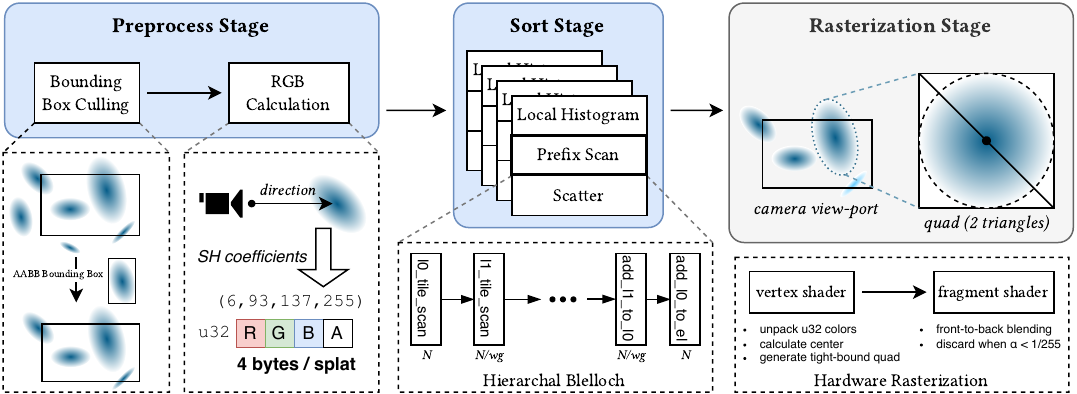}
\caption{System overview of \ours. The architecture follows a \textbf{hybrid compute-render pipeline}: blue stages represent compute passes, and gray stages represent render passes.}
\label{fig:pipeline}
\end{figure*}

\subsection{3D Gaussian Splatting}

The field of novel view synthesis has seen a progression from traditional explicit representations like meshes to implicit neural models such as Neural Radiance Fields (NeRF)~\cite{muller2022instant}. While NeRF offers high quality, its reliance on computationally intensive ray sampling and MLP queries limits real-time applications. 3D Gaussian Splatting (3DGS)~\cite{kerbl20233d} introduces an alternative by employing an explicit representation composed of 3D splats. This approach establishes a differentiable rendering pipeline, achieving both high-quality rendering and superior performance. However, this method introduces its own computational demands. To ensure correct rendering order, all Gaussians must be sorted by their distance from the camera. Similarly, simulating realistic lighting and reflections depends on view-dependent color calculations.

As an explicit 3D representation, 3DGS models a scene using a collection of 3D Gaussians, each defined by a set of geometric and appearance properties. The geometry distribution of each Gaussian is specified by its center position $X$ and a 3D covariance matrix $\Sigma$:
$$
G(\mathbf{x}) = e^{-\frac{1}{2}(\mathbf{x}-X)^T \Sigma^{-1}(\mathbf{x}-X)}.
$$

The covariance matrix $\Sigma$ is decomposed into a rotation quaternion and a scaling vector to define the ellipsoid's orientation and extent. The appearance of each Gaussian is captured by an opacity attribute $\sigma$ and view-dependent color, represented by Spherical Harmonics (SH) coefficients. To render an image from a given camera viewpoint, the 3D Gaussians are first projected onto the 2D image plane. The splatting opacity $\alpha$ for each Gaussian's contribution to a pixel is calculated based on its learned peak opacity $\sigma$ and the projected 2D covariance matrix $\Sigma'$:
$$
\alpha = \sigma e^{-\frac{1}{2}\mathbf{x'}^T \Sigma'^{-1}\mathbf{x'}},
$$
where $\mathbf{x'}$ is the distance vector from the pixel center to the projected Gaussian center. The color $c_i$ of each Gaussian is determined by evaluating its SH coefficients using the viewing direction $\mathbf{d}$ from the camera to the Gaussian's center:
$$
c_i = \text{SH}(k_i, \mathbf{d})
$$
Here, $\text{SH}$ represents the evaluation of the Spherical Harmonics basis functions with the learned coefficients $k_i$ for the specific camera view direction $\mathbf{d}$. This enables the Gaussian Splats to simulate reflection effects.

The final color $C$ for each pixel is then computed by alpha-blending the $N$ contributing Gaussians, which are sorted by depth to the camera from back to front:
$$
C = \sum_{i \in N} c_i \alpha_i \prod_{j=1}^{i-1} (1 - \alpha_j).
$$

\subsection{Web Graphics Rendering API}

Introduced by the Khronos Group in 2011, WebGL~\cite{webgl1} is a graphics API for web browsers, derived from OpenGL ES~\cite{gles}. WebGL's programming model faces significant performance limitations on modern hardware~\cite{han_gl2gpu_2025}. To further leverage modern GPUs, WebGPU was officially introduced in 2021~\cite{webgpu} as the next-generation graphics API for web browsers, offering a declarative programming model with more fine-grained resource management. The programming model of WebGPU is inspired by modern native APIs such as Vulkan, Metal, and DirectX~12.

\textbf{WebGPU Rendering Process.} The rendering process in WebGPU involves defining all necessary resources through descriptors. These descriptors are then used to construct WebGPU resources, which are filled in the GPU command buffers. A central WebGPU resource is the \texttt{Pipeline}, which encapsulates the vertex and fragment shaders. The vertex shader is responsible for computing vertex positions and organizing them into primitives like triangles. The fragment shader processes each pixel, interpolating vectors passed from the vertex shader (such as color or the distance vector $\mathbf{x}$ within a Gaussian splat). The \texttt{BindGroup} defines the layout for rendering data, such as RGB colors, positions, or splat centers. These individual resources are assembled into a \texttt{RenderPass}. Finally, a command encoder compiles the render pass into a command buffer, which is submitted to the GPU for execution. A detailed overview of the WebGPU rendering workflow is provided in Appendix~\ref{sec:webgpu_workflow}.

\textbf{WebGPU Compute Shaders.} One of the pivotal advancements of WebGPU over WebGL is its native support for general-purpose computation via \textit{compute shaders}. Since WebGL does not support general GPU computation, complex workloads like sorting or physics simulation must run on the CPU (through JavaScript or WebAssembly), leading to high data transfer overheads. In contrast, WebGPU's compute shaders provide direct access to the GPU's Single-Instruction-Multiple-Thread (SIMT) programming architecture. They execute on a grid of workgroups, where each workgroup comprises multiple threads that run in parallel and can communicate efficiently using shared memory. This powerful computational model is essential for modern rendering techniques that require massive data processing before rasterization, such as the GPU-based sorting algorithm and color calculation.
\section{Design}\label{sec:design}

This section introduces the design of \ours, including an overview of its workflow and details of the modules.

\subsection{Overview}

As illustrated in Figure~\ref{fig:pipeline}, the architecture of \ours is designed as a hybrid compute-render pipeline. This design fully leverages the capabilities of heterogeneous GPUs while maintaining broad hardware compatibility.

The pipeline consists of three stages. The \textit{Pre-processing Stage} uses compute shaders for view-dependent culling, 2D projection, and color computation of visible splats. The \textit{Sort Stage} performs a GPU-based radix sort on view-space depth using a wait-free algorithm that remains efficient across different GPU architectures. The \textit{Rasterization Stage} employs the hardware rendering pipeline to draw 2D Gaussians, reducing overhead by estimating each splat’s bounding box from its opacity to minimize redundant fragment processing.

\subsection{Pre-processing Stage}

The goal of the Pre-processing Stage is to transform the raw 3D Gaussian data into a compacted, view-dependent set of 2D primitives ready for sorting and rasterization. As shown in Figure~\ref{fig:pipeline}, this stage is composed of Bounding Box Culling and View-dependent Color Calculation.

\textbf{Bounding Box Culling.} For each Gaussian, we project its 3D center and covariance matrix into 2D screen space. This projection yields a 2D ellipse, for which we compute an axis-aligned bounding box (AABB). We drop the Gaussian whose bounding box does not intersect the camera viewport. The alive primitives are then written into a buffer for the subsequent stages. During this culling, \ours also calculates the major and minor axes of its projected 2D ellipse. We then compress and pack these two vectors into a pair of 32-bit unsigned integers (using 16-bit floats for each vector component) and pass them to the Rasterization Stage. We also calculate the depth for each primitive during the culling pass and store this depth as a 32-bit floating point into a dedicated buffer.

\textbf{View-Dependent Color Calculation.} For each splat, we evaluate its Spherical Harmonics (SH) coefficients based on the current viewing direction to compute the final RGB color. The resulting floating-point color and the splat's opacity are then compressed and packed into a single 32-bit unsigned integer using an \texttt{RGBA8} format. This data compression significantly reduces the memory bandwidth required during the final rendering stage.

\subsection{Sort Stage}

Following the pre-processing stage, \ours generates a list of visible splats. These splats must be rendered in back-to-front order to achieve correct alpha blending.

\subsubsection{The Challenge of Sorting on WebGPU}

In parallel radix sort, the computational bottleneck often lies in the efficiency of the prefix-sum (scan) primitive, which is used to calculate the starting offset for each element in the output array in parallel~\cite{sort1,sort2,sort3}. However, directly porting native GPU implementations is impractical because the WebGPU standard, designed for cross-platform compatibility, provides no guarantee on workgroup scheduling order or the fine-grained behavior of atomic operations. As a result, algorithms that rely on specific scheduling assumptions may perform inconsistently across devices.

A typical example is the inter-workgroup spin-wait pattern, where each workgroup computes local sums and then continuously polls a global atomic flag in a \texttt{while(true)} loop, waiting for other workgroups to finish. This approach implicitly assumes a fixed scheduling order; on devices without such guarantees, it degenerates into a busy-wait, wasting GPU cycles and memory bandwidth.

\subsubsection{A Wait-Free Radix Sort}

We design our sorting algorithm based on a wait-free, hierarchical Blelloch scan. 

\textbf{Design.} The core principle of the algorithm is that the final position (denoted as $pos$) for an element originally at index \texttt{i} during each sorting pass is determined by the formula:
$$pos = \text{base}_d + \text{rank}_i$$
In this formula, $d$ represents the specific digit value of the element being processed during the current pass of the radix sort. $\text{base}_d$ is a global offset for a given digit value, which is equal to the total count of all elements with a digit value less than $d$. $\text{rank}_i$ is defined as the count of elements that appear before index $i$ in the input array and share the same digit value $d$. This calculation of $pos$ guarantees that elements with identical digits are placed into the output array in the same relative order in which they appeared in the input.

To compute $\text{rank}_i$ efficiently, we decompose its calculation into two parts: an inter-workgroup component and an intra-workgroup component. This decomposition is the key to our parallel framework:
$$
\text{rank}_i=\text{prefix}_{wg}+\text{rank\_local}_i
$$

Where $\text{prefix}_{wg}$ is the inter-workgroup rank, representing the total count of elements with digit $d$ that appear in all workgroups processed before the current one; and $\text{rank\_local}_i$ is the intra-workgroup rank, representing the number of elements with digit $d$ that appear within the current workgroup but before $i$. Both the $\text{base}_d$ and $\text{prefix}_{wg}$ values are outputs of a global prefix sum.

\textbf{Implementation.} We apply our sorting design to the depth of each splat. We perform the full 32-bit sort by executing four consecutive passes of an 8-bit radix sort with a radix size of $256$. Starting from the least significant bits, each pass stably sorts the keys by an 8-bit segment of the depth value. Specifically, the $i$-th pass processes bits $i \times 8$ to $(i+1) \times 8 - 1$ of the key. Each of the four passes is composed of a three-step pipeline using WebGPU compute shaders. In the first step, \textit{Local Histogram}, each workgroup counts digit frequencies within each workgroup. In the second step, \textit{Prefix Scan}, the hierarchical Blelloch scan processes the histograms to compute the two required offsets for the scatter phase: the inter-workgroup prefix, $\text{prefix}_{wg}$, and the global $\text{base}_d$. In the third step, \textit{Scatter}, the algorithm writes each element to its new position $pos$ in the sorted array.

As shown in the Sort Stage in Figure~\ref{fig:pipeline}, we compute the per-digit offsets with a two-phase hierarchical Blelloch scan. In the \textit{forward pass} (upsweep), level~$L_0$ performs an exclusive scan inside each workgroup tile of size \texttt{wg} over the indicator stream for digit $d$. This produces tile sums, which form an array of length $N/\texttt{wg}$. Level~$L_1$ runs the same Blelloch scan on this array to obtain prefixes across tiles, and the process recurses until a single level remains. The last level provides global counts for all digits; from these, we read $\text{base}_d$ as the sum of counts of digits smaller than $d$. In the \textit{backward pass} (downsweep), the prefixes computed at level~$L_1$ are added back to their corresponding $L_0$ tiles, and then to individual elements, yielding the inter-workgroup offset $\text{prefix}_{wg}$. Each level is executed as an independent compute dispatch that consumes the previous level’s buffers, with only intra-workgroup barriers inside a tile. There is no cross-workgroup polling, so the algorithm makes linear work $O(N)$, uses $O(N/\texttt{wg})$ auxiliary storage, and avoids spin waiting across heterogeneous GPU schedulers.

\subsection{Rasterization Stage}

The final stage of our pipeline is responsible for rendering the sorted 2D Gaussians into the final image.

A key design choice in \ours is to use the hardware-accelerated vertex and fragment shader pipeline instead of a tile-based rasterizer. Tile-based methods, common in CUDA implementations, build and process a large buffer of $\langle tile_i, splat_i\rangle$ pairs that map screen tiles to intersecting splats. Generating and sorting this buffer incurs heavy memory traffic. On heterogeneous devices such as laptops and mobile phones, limited bandwidth makes this step a major bottleneck. By relying on the standard rendering pipeline, \ours avoids this cost and achieves better scalability in web environments.

As shown in Figure~\ref{fig:pipeline}, we render each Gaussian by submitting a single quad (composed of two triangles) via an instanced draw call.
Since each Gaussian represents a 3D ellipse with a spatial extent that projects differently depending on the camera view, a key challenge is to determine the appropriate screen-space size for each quad.
Ideally, the quad's bounds should tightly bound the area where the Gaussian's color and alpha contribution are significant. \ours derives this size from the Gaussian's opacity. The alpha contribution of a 2D Gaussian at a squared distance $r^2$ from its center is proportional to $e^{-r^2}$. The final opacity of a fragment is the product of this Gaussian decay and the splat's peak opacity, $\sigma$. We define the cutoff radius as the point where this final opacity drops below a minimum threshold, such as $1/255$:
$$\frac{1}{255} = \sigma \cdot e^{-r^2}$$
Solving for the radius $r$ gives the required scaling factor for the quad's vertices:
$$r = \sqrt{\ln(255 \cdot \sigma)}$$
This $r$ is computed in the vertex shader to size each quad, ensuring a tight bound dynamically. Gaussians with $\sigma < 1/255$ are culled in the pre-processing stage, as their contribution is below the quantization threshold.

A key insight in our rasterization is to evaluate the Gaussian function efficiently by reusing the hardware’s linear interpolators delivered by the render pipeline. As shown in~\ref{fig:pipeline}, the vertex shader passes a \textit{local coordinate} for each of the four quad vertices to the fragment shader. During rasterization, the hardware interpolates this local coordinate for each fragment, assigning each fragment its own position relative to the Gaussian center. For a fragment with barycentric coordinates $(\alpha, \beta, \gamma)$ within a triangle defined by vertices $(v_0, v_1, v_2)$, its interpolated local coordinate $p$ is:
$$p = \alpha v_0 + \beta v_1 + \gamma v_2$$

In the fragment shader, we then compute the dot product of this interpolated local coordinate with itself to get the squared distance from the center, denoted as $a$. This value $a$ is used directly in the Gaussian function $e^{-a}$ to calculate the final color and alpha contribution for that specific pixel, reconstructing the splat efficiently.

\section{Evaluation}\label{sec:evaluation}



\subsection{Experimental Setup}\label{sec:exp_setup}

We implement \ours in TypeScript and WGSL with approximately 6,600 lines of code. We employ Vite~\cite{vite} to compile the source code. \ours reads and renders Gaussian splatting scenes from standard \texttt{PLY}~\cite{Turk1994PLY} files.

\textbf{Devices.} We select a diverse range of devices to ensure our evaluation is representative of the heterogeneous hardware landscape of the modern web ecosystem.

\begin{itemize}[leftmargin=1.5em]
\item \textit{Desktop.} We use a high-end workstation with an Nvidia RTX 3070 GPU to establish the upper performance bound, and an Intel NUC mini-PC with an integrated Core i9-9980HK GPU to represent systems without discrete graphics.

\item \textit{Laptops.} We select laptops from both macOS and Windows ecosystems. We use a MacBook Pro (M1) and a MacBook Air (M4) to cover two generations of Apple Silicon. We also use a Lenovo laptop with an entry-level Nvidia MX350 GPU, representing typical thin-and-light Windows notebooks.

\item \textit{Mobile Devices.} We utilize three smartphones: an iPhone 15 Pro Max, a Redmi K70 Pro (Snapdragon 8 Gen 3), and an Oppo Find X2 (Snapdragon 865), representing current and previous hardware generations.
\end{itemize}

\textbf{Browsers.} We evaluate browsers with WebGPU available out of the box. Google Chrome has shipped WebGPU by default since version 113, released in April 2023~\cite{chrome_webgpu_overview}. In September 2025, Apple officially introduced native WebGPU support in Safari on iOS 26 and macOS 26, covering both mobile and desktop platforms~\cite{safari26_notes,webkit_safari26}. Firefox provides WebGPU support starting from version 141, currently limited to Windows platforms~\cite{moz_firefox141_webgpu,mdn_webgpu}. Therefore, our evaluation is primarily conducted on Chrome. All experiments are performed with Vertical Synchronization (V-Sync) enabled to avoid tearing and ensure consistent frame presentation.

\textbf{Baselines.} Web-based 3DGS currently lacks published academic baselines; existing works are community implementations derived from the original 3DGS paper. These implementations differ in engineering choices such as asynchronous scheduling or parameter packing. Therefore, we select the most widely adopted and actively maintained open-source viewers that represent the de facto state-of-the-art on the Web. For WebGL, we choose \texttt{antimatter15/splat} (\textbf{S1}) and \texttt{mkkellogg/GaussianSplats3D} (\textbf{S2}). For WebGPU, we select \texttt{KeKsBoTer/web-splat} (\textbf{W1}) and \texttt{MarcusAndreasSvensson/ gaussian-splatting-webgpu} (\textbf{W2}). As of October 2025, these represent the dominant approaches in the ecosystem.

\textbf{Benchmark Scenes.} Following previous work~\cite{kerbl20233d,radl_stopthepop_2024,lu_scaffold-gs_2024,feng_flashgs_2025,wang_adr-gaussian_2024}, we evaluate \ours's performance on a set of commonly used benchmark scenes to ensure a fair and comprehensive comparison. The scenes include \textit{bicycle} (6,131,954 splats), \textit{garden} (5,834,784 splats), \textit{truck} (2,541,226 splats), \textit{bonsai} (1,244,819 splats), \textit{train} (1,026,508 splats), and a cleaned version of the bicycle scene, \textit{bicycle-c} (1,063,091 splats). In addition to this collection, we also include the \textit{Van Gogh Room} (341,294 splats), a synthetic indoor scene from the Habitat test suite~\cite{habitat19iccv}. This selection of scenes provides a wide range of complexity, with splat counts spanning from approximately 340,000 to over 6 million, allowing us to evaluate the scalability of \ours thoroughly.

\begin{figure*}[t!]
\centering
\includegraphics[width=\textwidth]{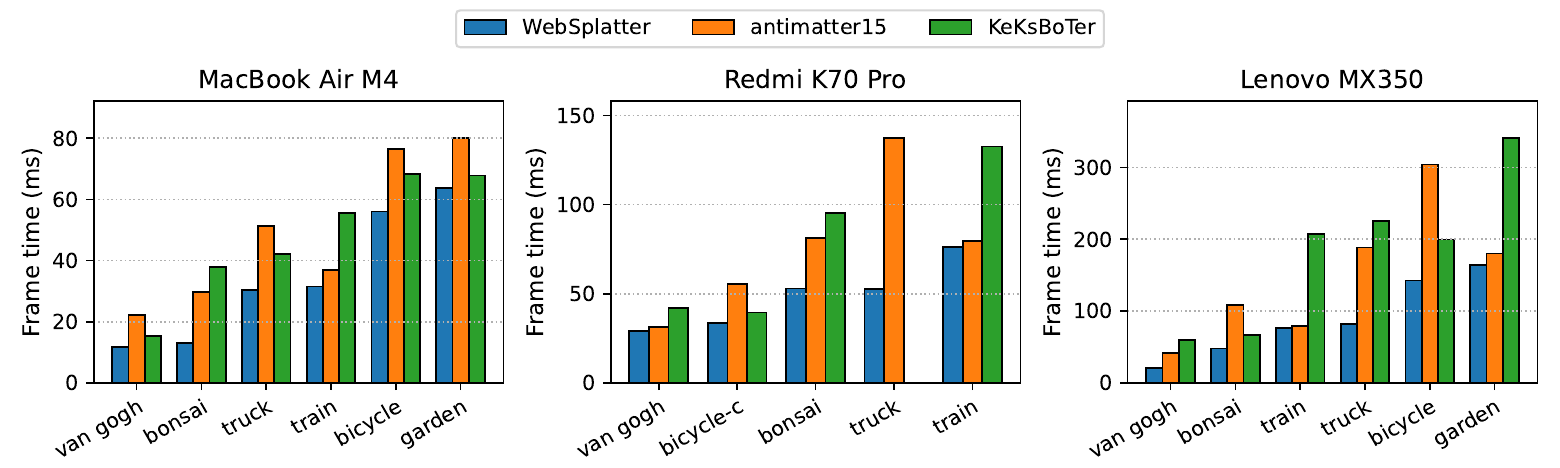}
\caption{Scalability comparison of \ours against baselines across three devices. Each bar group shows the median frame time (ms) for a specific scene. Lower is better. The \texttt{KeKsBoTer/web-splat} baseline crashes when rendering the \textit{truck} scene on the Redmi K70 Pro, and its result is omitted.}
\label{fig:scalability_chart}
\end{figure*}

\subsection{Cross-Platform Performance}

In this section, we present the evaluation results to answer three research questions. We begin with an overall performance comparison, followed by an analysis of scalability.

\begin{table}[t!]
\centering
\caption{Overall performance comparison showing total frame time in milliseconds (ms). Lower is better. Results are collected in \textbf{Chrome} unless annotated; \textit{(S)} indicates Safari and \textit{(F)} indicates Firefox. A dashed line indicates a crash. Abbreviations are defined in \S~\ref{sec:exp_setup}.}
\label{tab:overall_total_time_compact}
\resizebox{\columnwidth}{!}{%
\begin{tabular}{lllrrrrr}
\toprule
\textbf{Device} & \textbf{Scene} & \textbf{Ours} & \textbf{S1} & \textbf{S2} & \textbf{W1} & \textbf{W2} \\
\midrule
RTX 3070                 & garden           & \textbf{9.5}   & 62.4  & 88.8  & 14.4  & 15.1  \\
RTX 3070 \textit{(F)}    & garden           & \textbf{11.4}   & 54.9  & 67.8  & \textemdash{}  & \textemdash{}  \\
MacBook Air M4           & garden           & \textbf{63.7}  & 80.1  & 144.3 & 67.8  & 268.3 \\
MacBook Air M4 \textit{(S)}       & garden           & \textbf{68.6}  & 78.5  & 111.6 & 81.3  & \textemdash{} \\
MacBook Pro M1           & garden           & \textbf{112.0} & 124.4 & 225.2 & 510.8 & 402.3 \\
Nvidia MX350             & garden           & \textbf{164.3} & 180.0 & 869.5 & 341.0 & \textemdash{} \\
Intel NUC (iGPU)         & garden           & \textbf{151.2} & 341.7 & 368.4 & 404.0 & \textemdash{} \\
iPhone 15 Pro Max \textit{(S)}    & bicycle-c        & \textbf{38.5}  & 50.5  & \textemdash{} & 45.7  & \textemdash{} \\
Redmi K70 Pro            & bicycle-c        & \textbf{33.6}  & 55.3  & 76.4  & 39.5  & \textemdash{} \\
Oppo Find X2             & bicycle-c        & \textbf{105.2} & 112.7 & 182.6 & \textemdash{} & 4553.6 \\
\bottomrule
\end{tabular}
}
\end{table}

The overall performance across the test devices is presented in Table~\ref{tab:overall_total_time_compact}. We report the median total frame time for each test, encompassing the pre-processing, sorting, and rendering stages. The results demonstrate that \ours consistently and outperforms all baselines on every device, from high-end desktops to mobile phones. On the high-performance desktop with an Nvidia RTX 3070, \ours achieves a frame time of 9.497 ms on the \textit{garden} scene, representing a $1.52\times$ speedup over the fastest competing WebGPU implementation. This performance advantage extends across all hardware specifications. For instance, \ours is $1.06\times$ faster on the MacBook Air M4, $1.18\times$ faster on the flagship Redmi K70 Pro mobile phone, and achieves its most substantial relative gain on the Intel NUC with integrated graphics, where it is $2.26\times$ faster than the best-performing baseline. Notably, the performance gap is even wider on the MacBook Pro M1. The \texttt{KeKsBoTer/web-splat} baseline, which uses a spin-wait-based sort ported from native Fuchsia, performs poorly on Apple M1 chips. The lack of guarantees for workgroup execution order on this architecture leads to significant busy-waiting, highlighting the importance of our wait-free design. In this case, our method is over $4.5\times$ faster than the WebGPU baseline.

To further evaluate the stability of our system and address the concern that average frame times may obscure latency spikes, we analyzed the statistical distribution of frame times on the MacBook Air M4 using the computationally demanding \textit{garden} scene ($5.8$M splats). \ours exhibits remarkable consistency, with a standard deviation of only \textbf{22.303} ms and a 99th-percentile (P99) latency of \textbf{116.851} ms. In contrast, the baseline methods show significant jitter, primarily due to garbage collection overheads in WebGL or synchronization stalls in naive WebGPU ports. This low variance confirms that our wait-free sorting mitigates micro-stuttering, ensuring a fluid interactive experience even under heavy loads.

Furthermore, we find that Firefox remains unstable within the current WebGPU ecosystem. On the RTX 3070, \ours runs in Firefox with a slightly higher frame time ($11.4$ ms), whereas both WebGPU baselines (W1 and W2) crash. On our Lenovo MX350 system, Firefox blocks WebGPU at startup with the error \emph{``WebGPU is disabled by blocklist''}. Meanwhile, on the Intel NUC device, WebGPU timestamp queries in Firefox appear unreliable: while the built-in queries report a frame time of $2.8$ ms, the JavaScript API measures $258.36$ ms. This discrepancy has been confirmed as a Firefox bug~\cite{timestamp_bug}. Due to these issues, we omit Firefox results for both the Lenovo MX350 and Intel NUC systems.

Beyond rendering speed, viewer availability is critical for web-based applications. \ours successfully renders all scenes on all test devices, demonstrating its wide compatibility. In contrast, several baselines fail on certain devices due to excessive memory requirements that surpass hardware limitations. For instance, the WebGPU-based \texttt{KeKsBoTer/web-splat} viewer crashes on the Oppo Find X2. Although its pre-processing and sorting stages are complete, the application fails upon entering the rendering stage. Similarly, the WebGL-based \texttt{mkkellogg/GaussianSplatsD} viewer repeatedly causes the browser to crash on the iPhone 15 Pro Max.

\begin{table}[t!]
\centering
\caption{Performance breakdown of \ours across representative devices. Results are collected in \textbf{Chrome} unless annotated; \textit{(F)} indicates Firefox. Total frame times are in milliseconds (ms). Percentages denote the fraction of time spent in each stage.}
\label{tab:breakdown_analysis}
\resizebox{\columnwidth}{!}{
\begin{tabular}{l l r r r r}
\toprule
\textbf{Device} & \textbf{Scene} & \textbf{Total} & \textbf{Pre-process} & \textbf{Sort} & \textbf{Render} \\
\midrule
RTX 3070           & garden     & 9.5   & 13.1\% & 24.3\% & \textbf{62.6\%} \\
RTX 3070 \textit{(F)} & garden     & 11.4  &  5.6\% & 37.7\% & \textbf{56.8\%} \\
MacBook Air M4     & garden     & 63.7  & 30.1\% & 18.1\% & \textbf{51.8\%} \\
MacBook Pro M1     & garden     & 112.0 & 24.3\% & 17.0\% & \textbf{58.7\%} \\
Nvidia MX350       & garden     & 164.3 & \textbf{47.2\%} & 27.6\% & 25.3\% \\
Intel NUC (iGPU)   & garden     & 151.2 & \textbf{40.0\%} & 29.2\% & 30.8\% \\
Redmi K70 Pro      & bicycle-c  & 33.6  & 10.2\% & 28.8\% & \textbf{61.0\%} \\
Oppo Find X2       & bicycle-c  & 105.2 & 13.4\% & 37.7\% & \textbf{48.8\%} \\
\bottomrule
\end{tabular}
}
\end{table}

\textbf{Bottleneck analysis.} We present an analysis of the bottleneck on different devices in Table~\ref{tab:breakdown_analysis}. On systems with high-end graphics capabilities (including Nvidia RTX 3070, Apple M-series, and even contemporary mobile Redmi K70 Pro), the system is primarily render-bound. The rasterization stage on these devices is the most expensive, accounting for 51.8\% to 62.6\% of the total frame time. In contrast, on systems with less powerful GPUs, such as the Nvidia MX350 and the Intel NUC's integrated graphics, the bottleneck shifts to the pre-processing stage, consuming up to 47.2\% of the frame time. This shift is because a large number of raw splats need to be scanned in the pre-processing stage. The limited memory throughput of these low-end devices causes performance degradation. Meanwhile, \ours never becomes sort-bound across all tested hardware. This result demonstrates the efficiency of our wait-free sorting algorithm.

\begin{figure*}[t]
\centering
\begin{minipage}{\columnwidth}
  \centering
  \includegraphics[width=\linewidth]{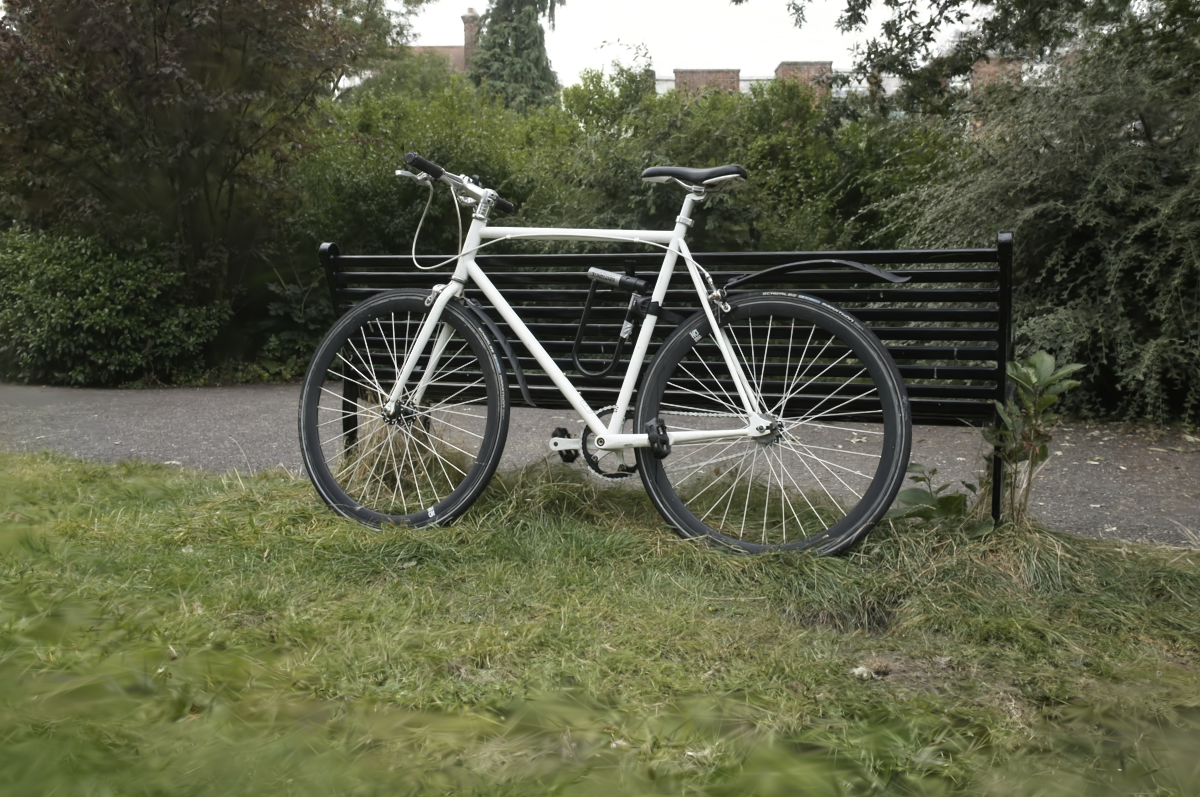}
\end{minipage}
\hfill
\begin{minipage}{\columnwidth}
  \centering
  \includegraphics[width=\linewidth]{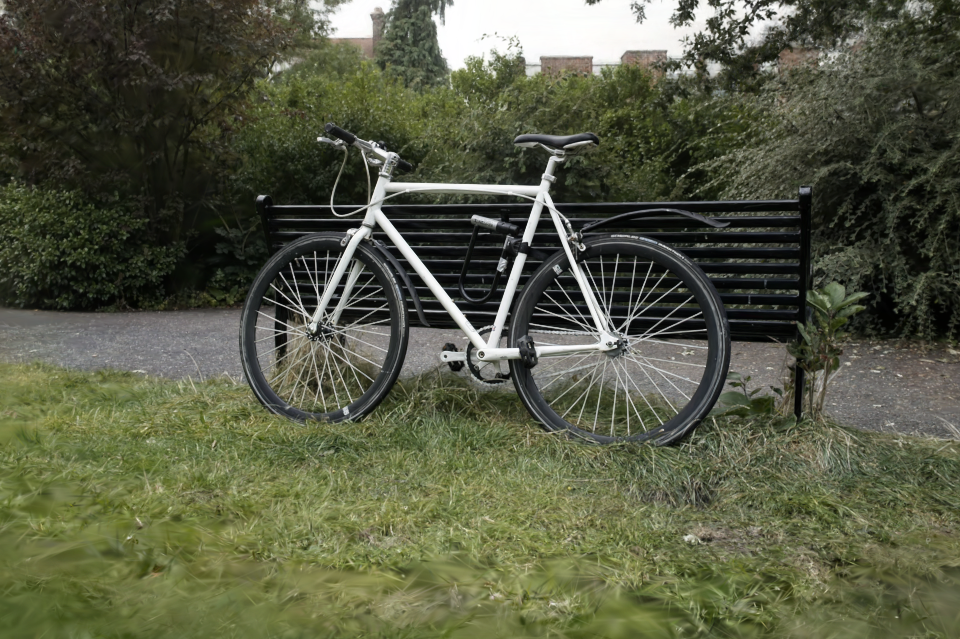}
\end{minipage}
\caption{Visual correctness verification on the \textit{garden} scene. \textbf{Left:} Output rendered by the official native CUDA implementation. \textbf{Right:} Output rendered by \ours in Google Chrome.}
\label{fig:visual_comparison}
\end{figure*}

\textbf{Rendering Quality.} In addition to performance, we also assess the visual fidelity of our renderer. Figure~\ref{fig:visual_comparison} presents a side-by-side comparison between the official native CUDA implementation and \ours. The results show that our WebGPU pipeline achieves a visual quality comparable to the native baseline, suggesting that our optimizations effectively maintain the rendering characteristics of the original 3DGS algorithm.

\subsection{Scalability with Scene Complexity.}

We evaluate how the performance of \ours scales with increasing scene complexity. We select three representative devices, exhibiting two different bottlenecks as identified in our overall analysis: the render-bound MacBook Air M4, the render-bound K70 Pro for mobile, and the pre-process-bound Nvidia MX350. On these platforms, we compare \ours against two key baselines: \texttt{KeKsBoTer/web-splat}, as the strongest WebGPU competitor, and \texttt{antimatter15/splat}, representing a traditional WebGL viewer with CPU-based sorting. We utilize our full set of benchmark scenes to offer a range of complexities. The number of splats (denoted as $N$) ranges from approximately 340,000 to over 6 million. For each scene on each device, we maintain a fixed resolution and camera position. We measure the median frame time and the breakdown of time for \ours and other baselines.

The overall scalability comparison is demonstrated in Figure~\ref{fig:scalability_chart}. The results show that the performance of \ours scales more effectively with increasing scene complexity compared to both the WebGPU and WebGL baselines across all three representative device tiers. For the pre-process-bound Nvidia MX350, as scene complexity increases from the \textit{van gogh} scene (0.3M splats) to the \textit{garden} scene (5.8M splats), the frame time for the \texttt{KeKsBoTer/web-splat} baseline increases by over 280 ms. In contrast, the frame time for \ours increases by a much smaller margin of approximately 143 ms, resulting in a $2.1\times$ speedup in the largest scene. Furthermore, our approach demonstrates greater stability. On the mobile Redmi K70 Pro, \ours renders the complex \textit{truck} scene (2.5M splats) in 52.7 ms, while the competing WebGPU viewer crashes under the same load. The \texttt{antimatter15} baseline exhibits a high frame time. Specifically, its rendering time rises to 188.4 ms on the Lenovo MX350 and $137.37$ ms on the Redmi K70 Pro, and $51.4$ ms on the MacBook Air M4.

\begin{table}[t!]
\centering
\caption{Median times (ms) across scenes with different splat counts, broken down by three pipeline stages.}
\label{tab:scalability_results_ours}
\resizebox{\columnwidth}{!}{%
\begin{tabular}{l r | rrrr}
\toprule
& & \multicolumn{4}{c}{\textbf{\ours (ms)}} \\
\textbf{Scene} & \textbf{Splats (N)} & \textbf{Total} & \textbf{Pre-proc} & \textbf{Sort} & \textbf{Render} \\
\midrule
\multicolumn{6}{l}{\textit{\textbf{Device: MacBook Air M4 (Render-Bound)}}} \\
\quad van gogh   & 341,294   & \textbf{11.654} & 0.761 & 2.046 & 8.847 \\
\quad train      & 1,026,508 & \textbf{31.518} & 5.332 & 8.522 & 17.664 \\
\quad bonsai     & 1,244,819 & \textbf{13.088} & 1.349 & 1.746 & 9.993 \\
\quad truck      & 2,541,226 & \textbf{30.350} & 7.715 & 6.805 & 15.830 \\
\quad bicycle    & 6,131,954 & \textbf{56.013} & 19.640 & 7.683 & 28.690 \\
\quad garden     & 5,834,784 & \textbf{63.737} & 19.183 & 11.518 & 33.036 \\
\midrule
\multicolumn{6}{l}{\textit{\textbf{Device: Redmi K70 Pro (Render-Bound)}}} \\
\quad van gogh   & 341,294   & \textbf{28.960} & 0.422 & 4.171 & 24.367 \\
\quad train      & 1,026,508 & \textbf{76.369} & 2.185 & 5.066 & 69.118 \\
\quad bicycle-c  & 1,063,091 & \textbf{33.633} & 3.431 & 9.673 & 20.529 \\
\quad bonsai     & 1,244,819 & \textbf{52.874} & 2.290 & 5.688 & 44.896 \\
\quad truck      & 2,541,226 & \textbf{52.719} & 2.101 & 4.435 & 46.183 \\
\midrule
\multicolumn{6}{l}{\textit{\textbf{Device: Lenovo MX350 (Pre-process-Bound)}}} \\
\quad van gogh   & 341,294   & \textbf{20.946}  & 1.360  & 9.104  & 10.482 \\
\quad train      & 1,026,508 & \textbf{76.791}  & 22.861 & 15.985 & 37.945 \\
\quad bonsai     & 1,244,819 & \textbf{47.665}  & 15.439 & 11.297 & 20.929 \\
\quad truck      & 2,541,226 & \textbf{81.666}  & 29.702 & 18.650 & 33.314 \\
\quad bicycle    & 6,131,954 & \textbf{142.771} & 67.336 & 38.659 & 36.776 \\
\quad garden     & 5,834,784 & \textbf{164.255} & 77.452 & 45.253 & 41.550 \\
\bottomrule
\end{tabular}
}
\end{table}

Furthermore, for \ours, the results reveal a consistent scaling trend as scene complexity increases. As shown in Table~\ref{tab:scalability_results_ours}, in low-complexity scenes such as \textit{van gogh} (0.34 M splats), the system remains render-bound across all devices. The rendering stage accounts for about 84\% of the total frame time on the Redmi K70 Pro (24.4 ms of 28.9 ms), 78\% on the MacBook Air M4 (8.8 ms of 11.7 ms), and roughly 50\% on the Lenovo MX350 (10.5 ms of 20.9 ms). As the splat count grows to a million-level, the pre-processing cost increases more rapidly than the rendering cost. This increase is particularly evident on the MX350: the pre-processing stage expands from 1.36 ms (7\%) in the \textit{van gogh} scene to 77.5 ms (47\%) in the \textit{garden} scene, where the total frame time reaches 164.3 ms. These results suggest that, for highly complex web scenes, optimizing data preparation and splats culling becomes as critical as optimizing the rasterization pipeline.

\subsection{Ablation Study}

To further underscore the contribution of our sorting design, we first present a direct comparison of the sort stage time against all baselines. As shown in Table~\ref{tab:sort_time_compact}, our sort stage consistently incurs lower overhead than all baselines across Chrome, Safari, and Firefox. The results highlight two key advantages of our approach. First, leveraging the hardware acceleration through WebGPU compute shaders, our GPU-based sorting is much more efficient than CPU-based sorting. On the Nvidia RTX 3070, our GPU-based radix sort is over $24\times$ faster than the CPU-based sort in \texttt{antimatter15/splat} (2.3 ms vs. 55.8 ms) and over $31\times$ faster than the WASM-based sort in \texttt{mkkellogg/GaussianSplats3D}. Second, our sorting algorithm is designed to achieve excellent performance without relying on the specific thread scheduling implementation of the underlying hardware. On the MacBook Pro M1, for instance, our sort is over $24\times$ faster than the spin-wait-based implementation in \texttt{KeKsBoTer/web-splat} (19.0 ms vs. 458.4 ms).

\begin{table}[t!]
\centering
\caption{Sort stage time (ms) across devices. Results are collected in \textbf{Chrome} unless annotated; \textit{(S)} indicates Safari and \textit{(F)} indicates Firefox. A dashed line indicates a crash. Abbreviations are defined in \S~\ref{sec:exp_setup}.}
\label{tab:sort_time_compact}
\resizebox{\columnwidth}{!}{%
\begin{tabular}{lrrrrr}
\toprule
\textbf{Device} & \textbf{Ours} & \textbf{S1} & \textbf{S2} & \textbf{W1} & \textbf{W2} \\
\midrule
RTX 3070                    & \textbf{2.306}  & 55.8   & 72.0   & 5.374   & 5.308  \\
RTX 3070 \textit{(F)}       & \textbf{4.307}  & 49.0   & 50.6   & \textemdash{}   & \textemdash{}  \\
Redmi K70 Pro               & \textbf{9.673}  & 39.7   & 48.6  & 17.629  & \textemdash{} \\
MacBook Air M4 \textit{(S)} & \textbf{15.022} & 32.0   & 48.6   & 20.557  & \textemdash{} \\
MacBook Pro M1              & \textbf{19.027} & 58.7   & 154.2   & 458.424 & 261.816 \\
Lenovo MX350                & \textbf{45.253} & 98.5   & 556.2 & 150.733 & \textemdash{} \\
\bottomrule
\end{tabular}
}
\end{table}

We now analyze the impact of our other major optimizations. To isolate the impact of each component, we start with our fully optimized pipeline (denoted as "FULL") and create separate configurations where each key optimization is disabled one at a time. We focus on two features:

To analyze the culling mechanism of our approach, we perform an ablation study on the remaining optimization components. Starting from the fully optimized pipeline (denoted as ``\textbf{FULL}''), we disable each component in turn to isolate its contribution.

\begin{itemize}[leftmargin=1.5em]
\item \textbf{-CULL:} We disable the screen-space AABB culling in this revision, retaining only the initial view frustum culling. As a result, the sort and render stages process a larger number of splats.

\item \textbf{-RADIUS:} We disable the opacity-based dynamic quad sizing in the vertex shader in this revision. All splats are rendered with a fixed-size bounding box.
\end{itemize}

\begin{table}[t!]
\centering
\caption{Ablation study results on the \textbf{MacBook Air M4} with the \textit{garden} scene. All timings are in milliseconds (ms).}
\label{tab:ablation_study}
\begin{tabular}{l r r r}
\toprule
\textbf{Pipeline Stage} & \textbf{FULL} & \textbf{-CULL} & \textbf{-RADIUS} \\
\midrule
Pre-process (ms) & 19.18 &  21.79 & 20.42 \\
Sort (ms)        & 11.52 &  11.05 & 13.59 \\
Render (ms)      & 33.04 &  34.52 & 39.20 \\
\midrule
\textbf{Total Time (ms)} & \textbf{63.74} & \textbf{67.36} & \textbf{73.20} \\
\bottomrule
\end{tabular}
\end{table}

We conduct this experiment on the MacBook Air M4, a representative render-bound platform. We use the \textit{garden} scene to ensure the performance impact of each optimization is measurable. As shown in Table~\ref{tab:ablation_study}, disabling the visibility culling pass (\textbf{-CULL}) has a more modest, though still clear, impact on this platform, increasing the total frame time by $5.8\%$ (from 63.7 ms to 67.4 ms). As predicted, the performance cost is primarily absorbed by the Pre-process stage. This is because the process stage must process all splats that pass the initial frustum cull.

Disabling the dynamic radius sizing (\textbf{-RADIUS}) has the most significant performance impact, increasing the total frame time by $15\%$ (from 63.7 ms to 73.2 ms). This performance loss is almost entirely isolated to the Render stage, whose workload increases by $18.8\%$ (from 33.0 ms to 39.2 ms). This is because rendering all splats with a larger, fixed-size quad leads to significant overdraw and increases the total number of pixels processed by the fragment shader. This demonstrates the effectiveness of our opacity-based tight bounding in optimizing pixel throughput. In summary, the ablation study confirms that both optimizations contribute positively to performance, with the tight bounding mechanism being particularly crucial for reducing rasterization costs.

\subsection{Memory Efficiency}

A critical factor for web-based deployment, particularly on mobile devices, is the consumption of video memory (VRAM). Excessive memory usage in existing WebGPU ports often results in browser context loss or crashes on lower-end hardware.

To quantify this, we measure the peak GPU memory usage on the \textit{garden} scene using an Nvidia RTX 3070 on Windows. \ours consumes approximately \textbf{1.20 GB} of VRAM, which is more efficient than competing WebGPU implementations: it reduces memory usage by $36\%$ compared to W1 (1.90 GB) and by over $57\%$ compared to W2 ($>2.80$ GB). Furthermore, our memory footprint remains comparable to the lightweight WebGL viewers S2 (1.19 GB) and S1 (0.83 GB), despite offering the high-performance benefits of a compute-driven pipeline. This efficient memory management explains the stability results observed in Table~\ref{tab:overall_total_time_compact}, where \ours successfully runs on memory-constrained mobile devices (e.g., iPhone 15 Pro Max, Oppo Find X2) while the unoptimized WebGPU baselines frequently crash due to out-of-memory errors.

\section{Related Work}\label{sec:relatedwork}


\subsection{The WebGPU API.}

In recent years, there has been a surge in research focusing on WebGPU, aiming for shader testing, security enhancements, and rendering performance improvements. Levine et al. introduced a technique for testing memory consistency in WebGPU Shader Language (WGSL)~\cite{10.1145/3575693.3575750,10.1145/3597926.3598095}. FusionRender~\cite{bi2024fusionrender} enhances end-to-end performance by merging object signatures in WebGPU. NNJit~\cite{jia2024nnjit} enables just-in-time (JIT) auto-generation of optimized WGSL kernels for edge devices. On the security front, Ferguson et al. explored cache attacks in WebGPU to identify browser clients through side-channel attacks~\cite{ferguson2024webgpuspy}, and Giner et al. employed similar techniques to conduct memory leakage attacks~\cite{giner2024generic}. GL2GPU~\cite{han_gl2gpu_2025} focuses on dynamically translating WebGL applications to WebGPU within the JavaScript runtime to boost end-to-end rendering performance. While many works study the API's capabilities and potential vulnerabilities, our work provides a practical design of WebGPU's advanced compute features to solve a demanding, real-world rendering challenge.

\subsection{3D Gaussian Splatting (3DGS) Acceleration.}

The original 3DGS~\cite{kerbl20233d} achieved an unprecedented balance of rendering quality and speed. Subsequent research has focused on further optimizing its performance, primarily through two avenues: model compression and rasterization pipeline enhancement.

\begin{itemize}[leftmargin=1.5em]

\item \textit{Model Compression.} Many work aims to reduce the model size or the memory footprint of 3DGS. These methods reduce the number of Gaussians through pruning strategies based on importance metrics~\cite{durvasula_distwar_2023, lee_compact_2024} or by modifying the densification process during training~\cite{girish2023eagles}. While effective at reducing model size, these approaches can sometimes compromise rendering quality and do not address the computation bottlenecks of the rendering itself.

\item \textit{Fast Rendering Pipeline.} Another line of research focuses on accelerating the rasterization pipeline, which is essential for handling large-scale scenes. AdR-Gaussian~\cite{wang_adr-gaussian_2024} and FlashGS~\cite{feng_flashgs_2025} identify that the initial Gaussian-tile intersection test is a major source of inefficiency. They propose using more precise intersection algorithms to reduce the number of redundant Gaussian-tile pairs that enter the sorting and rendering stages. StopThePop~\cite{radl_stopthepop_2024} addresses temporal flickering artifacts by introducing a more view-consistent sorting key, highlighting the critical role of the sorting stage. These methods, however, are typically implemented in CUDA for native applications. In contrast, our work focuses on redesigning the core components of the WebGPU pipeline in a cross-platform manner.
\end{itemize}

\subsection{Web-Based 3DGS Viewers.}

The accessibility of the web has led to the development of several 3DGS viewers~\cite{zhu_rendering_2025, sinha_spectralsplatsviewer_2024}. Early implementations primarily relied on WebGL. Due to WebGL's lack of compute shaders, computationally intensive tasks like per-frame depth sorting must be offloaded to the CPU (via JavaScript or WebAssembly), creating a significant bottleneck that prevents real-time rendering of large-scale scenes. Some web viewers focus on reducing network transmission overhead by employing quantization techniques to compress the Gaussian data~\cite{niedermayr2024compressed}. However, these efforts primarily address data loading rather than the core rendering performance. Existing web-based viewers have not fully leveraged the capabilities of modern web graphics APIs to tackle the challenges of the 3DGS pipeline.

\section{Conclusion}\label{sec:conclusion}

In this paper, we have presented \ours, a WebGPU-based framework that leverages advanced compute shaders to enable high-performance, fully GPU-driven 3D Gaussian Splatting in web browsers. Through a hybrid compute-render pipeline and a wait-free sorting algorithm, \ours have addressed web-specific challenges in WebGPU and achieved stable and efficient performance across heterogeneous hardware. Compared with the state-of-the-art baselines, our evaluation shows speedups ranging from $1.06$ to $2.26$ times across devices from high-end desktops to mobile phones. These results demonstrate that \ours effectively harnesses the computational power of WebGPU to deliver real-time, high-fidelity rendering while substantially improving cross-platform performance within the modern web graphics ecosystem.

\bibliographystyle{ACM-Reference-Format}
\bibliography{base}

\appendix

\section{Rendering Workflow of WebGPU}\label{sec:webgpu_workflow}

The design of WebGPU emphasizes explicit resource management and the pre-compilation of state to minimize CPU overhead during the per-frame render loop. The entire process is asynchronous and revolves around creating command buffers that are submitted to the GPU for execution. The following steps outline this workflow in detail.

\textbf{Initialization and Device Acquisition}
The entry point to the API is the asynchronous request for a \texttt{GPUAdapter}, which represents a specific physical GPU on the system. From this adapter, a \texttt{GPUDevice} is requested, also asynchronously. The \texttt{GPUDevice} is the primary interface for creating all WebGPU resources, such as buffers, textures, and pipelines. Associated with the device is a \texttt{GPUQueue}, which is the channel for submitting commands to the GPU for execution.

\textbf{One-Time Setup Resources and Pipelines}
Before rendering can occur, all necessary state and data must be defined and, where possible, uploaded to the GPU. This setup phase is typically performed once at application load time.

First, data such as vertex coordinates, normals, or uniform variables are stored in \texttt{GPUBuffer} objects, created via \texttt{createBuffer()}. A descriptor specifies the buffer's size and usage flags, which inform the driver how to optimize memory management.

Second, the core of the rendering state is encapsulated in a \texttt{GPURenderPipeline}. Its creation is a significant step where the GPU driver pre-compiles and validates the entire rendering configuration for maximum performance. This involves creating a detailed descriptor that specifies:
\begin{itemize}[leftmargin=1.5em]
    \item The vertex and fragment shaders, provided as a \texttt{GPUShaderModule} containing WGSL code.
    \item The layout of vertex data, including the stride of the vertex buffer and the format of each attribute.
    \item The format of the render target (e.g., the swap chain texture format) and depth/stencil buffer.
    \item The layout of bind groups (\texttt{GPUBindGroupLayout}), which defines the interface for connecting data like uniforms and textures to the shaders.
\end{itemize}

\textbf{Data Binding}
WebGPU uses a two-part system to bind resources to shaders. The \texttt{GPUBindGroupLayout} defines a ``template'' or interface for a group of resources, specifying their type and binding point within the shader (e.g., `@binding(0) uniform Params'). This layout is part of the pre-compiled pipeline. The actual resources (e.g., a specific \texttt{GPUBuffer}) are then linked to this template by creating a \texttt{GPUBindGroup} instance. This separation allows efficient switching of data sources at render time without re-validating the pipeline.

\textbf{Command Encoding}
The work for each frame is recorded into a command buffer. This process begins by creating a \texttt{GPUCommandEncoder}. To perform graphics rendering, a \texttt{GPURenderPassEncoder} is started from the command encoder via \texttt{beginRenderPass()}. This encoder requires a descriptor specifying the color attachments (the texture to draw on) and depth attachments.

Inside the render pass, a sequence of commands is recorded to instruct the GPU what to draw:
\begin{enumerate}[leftmargin=1.5em]
    \item \texttt{passEncoder.setPipeline(myPipeline)}: Activates the pre-compiled render pipeline.
    \item \texttt{passEncoder.setVertexBuffer(slot, myBuffer)}: Binds a vertex buffer to a specific input slot.
    \item \texttt{passEncoder.setBindGroup(index, myBindGroup)}: Binds a group of resources.
    \item \texttt{passEncoder.draw(vertexCount, instanceCount)}: Issues a draw call, which triggers the execution of the vertex and fragment shaders for the specified number of vertices and instances.
\end{enumerate}
Once all draw calls are recorded, the render pass is concluded with \texttt{passEncoder.end()}.

\textbf{Submission and Execution}
After all render passes and compute passes are recorded, the command encoder is finalized using \texttt{encoder.finish()}, which returns an immutable \texttt{GPUCommandBuffer}. This buffer is then submitted to the device's \texttt{GPUQueue}. The submission is non-blocking, meaning the CPU is immediately free to continue processing the next frame while the GPU works asynchronously to execute the submitted commands.



\section{Introduction to the Baselines}\label{sec:appendix_baselines}

As baselines, we select the two most-starred open-source Gaussian splatting projects for each of the two primary web graphics APIs: WebGL and WebGPU. To ensure the reproducibility of our experiments, we benchmark against specific commit versions of each project as detailed below.

\begin{itemize}[leftmargin=1.5em]
    \item \textbf{\texttt{antimatter15/splat}}: A minimalist WebGL implementation with no external dependencies that performs sorting on the CPU via a web worker. The version used in our evaluation is based on commit \texttt{367a943} from \url{https://github.com/antimatter15/splat}.

    \item \textbf{\texttt{mkkellogg/GaussianSplats3D}}: A feature-rich viewer built on Three.js that is optimized with WASM-powered sorting and octree-based culling. The version used is based on commit \texttt{2dfc83e} from \url{https://github.com/mkkellogg/GaussianSplats3D}.

    \item \textbf{\texttt{KeKsBoTer/web-splat}}: A high-performance WebGPU renderer built with Rust featuring an efficient GPU-based radix sort. The version used is based on commit \texttt{959a3ec} from \url{https://github.com/KeKsBoTer/web-splat}.
    
    \item \textbf{\texttt{MarcusAndreasSvensson/gaussian-splatting-webgpu}}: A faithful WebGPU implementation of the original paper's methodology. The version used is based on commit \texttt{f22ae04} from \url{https://github.com/MarcusAndreasSvensson/gaussian-splatting-webgpu}.
\end{itemize}

To ensure a fair and accurate comparison of rendering performance, we instrumented each baseline framework to measure GPU execution time precisely. This process isolates the time spent on GPU workloads from other factors such as CPU-side logic or JavaScript framework overhead. Our modifications were minimal and targeted only at performance profiling, using the standard APIs available for each graphics backend.

\end{document}